\documentclass[aps,prl,twocolumn,floatfix,superscriptaddress]{revtex4}
\usepackage{graphicx}
\begin{document}

\title{Broken time-reversal symmetry in  strongly correlated ladder structures}

\author{ U. Schollw\"ock}
\affiliation{Max Planck Institute for Solid State Research, Heisenbergstr. 1,
D-70569 Stuttgart, Germany}
\author{Sudip Chakravarty}
\affiliation{Department of Physics and Astronomy, University of California Los Angeles, CA 90095,
USA}
\author{J. O. Fj{\ae}restad}
\affiliation{Department of Physics, Brown University, Providence, RI 02912, USA}
\author{J. B. Marston}
\affiliation{Department of Physics, Brown University, Providence, RI 02912, USA}
\author{ M. Troyer}
\affiliation{Theoretische Physik, Eidgen\"ossische Technische Hochschule, CH-8093
     Z\"urich, Switzerland}
\date{\today}  

\begin{abstract}
We provide, for the first time, in a doped strongly 
correlated system (two-leg ladder), a controlled  theoretical demonstration of the existence 
of a state  in which long-range ordered orbital currents are arranged in a staggered pattern,
coexisting with a charge density wave. 
The method used is the highly accurate density matrix renormalization group technique.
This brings us closer to recent proposals that this order is realized in the
enigmatic pseudogap phase of the cuprate high temperature superconductors. 
\end{abstract}
\maketitle
The circulating current phases in correlated electron systems, also called orbital antiferromagnets (OAF), 
were first considered in 
the context of excitonic insulators \cite{orbitalafm}, but then discarded in favor of more conventional order. 
After the discovery of the cuprate high temperature superconductors
they were rediscovered \cite{flux,Schulz} and called staggered flux (SF) phases \cite{flux}. 
Many of their properties were discussed, but were forgotten again in the absence of experimental
vindication. The discovery of an unusual and robust regime called the pseudogap \cite{Timusk} 
in these superconductors has changed the picture once more. The pseudogap
mimics the momentum dependence of the superconducting gap, 
but the state itself  is not superconducting. In this context, two recent
developments have taken place: (1) attempts have been made to explain this regime in terms of 
fluctuations of SF order \cite{Patrick} and  (2) a proposal has been made that it
is not fluctuations, but a true broken symmetry that is the origin of the pseudogap \cite{ddw,Kee}. 
This ordered state was called the singlet $d$-density wave (DDW) following
Ref. ~\cite{Nayak} and is the same as the OAF and SF phases. In this Letter we adopt the density 
wave (DW) terminology, as it can describe large classes of order parameters
with orbital angular momentum. The label $d$ stands for angular momentum 2, as in atomic physics. 
The conventional charge density wave (CDW) in which charge is
modulated in space is its $s$-wave counterpart with angular momentum zero. 
The triplet $s$-wave density wave is what is commonly called a spin density wave (SDW).
Another breakdown of time reversal symmetry in which the circulating currents obey translational symmetry, 
as opposed to DDW, has also been pointed out and has
been argued to be responsible for the pseudogap phase \cite{Varma,Kaminski}.

\begin{figure}
\centerline{\includegraphics[scale=0.5]{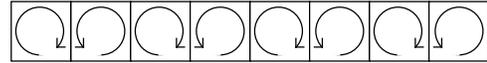}}
\vspace{0.5 cm}
\caption{Circulating plaquette currents on a ladder, characteristic of the DDW
phase.}
\label{fig:cur}
\end{figure}

Although much indirect experimental evidence of DDW can be argued to exist, 
a direct observation of DDW would be Bragg reflection of neutrons carrying 
magnetic moments from the staggered arrangement, on the scale of a few {\AA}, of
circulating currents. Recent neutron scattering experiments have, however, been 
controversial \cite{Mook,CKN,Sidis,Stock} 
and more precise and well-characterized experiments are underway
to establish this order. Thus, theoretical exploration of microscopic models of correlated electronic systems 
with controlled methods has acquired urgency. We,
therefore, study the simplest geometrical structure in the form of a two-leg ladder \cite{Dagotto}
that can support staggered orbital currents, as shown in Fig.~\ref{fig:cur}. 

Previous studies of DDW order in two-leg ladders have used weak-coupling 
bosonization/renormalization group (RG) analyses \cite{Nersesyan,Luther,Schulz2,
Orignac,Scalapino,John,Wu,Furusaki}, density-matrix renormalization group
(DMRG) \cite{DMRG} analysis of the $t$-$J$-model  \cite{Scalapino} and a half-filled Hubbard-like
model \cite{MFS},  or exact diagonalization \cite{Tsutsui}
of the $t$-$t'$-$J$-model. At half-filling, models with long-range ordered currents 
have been found both for spinless \cite{Nersesyan} and spinful
\cite{John,MFS,Wu,Furusaki} fermions. In contrast, for doped ladders, 
power-law decay has been found for
spinless \cite{Luther,Orignac} and spinful \cite{Schulz2,Orignac,Wu}
cases. For the $t$-$J$ ladder, only short-range order was found
\cite{Scalapino}, and the study of the $t$-$t'$-$J$ model \cite{Tsutsui} yielded similar results. 

The approach used in the present work is the accurate DMRG method 
that can be used for arbitrary interaction strength, 
unlike the weak-coupling bosonization/RG approaches. 
The results of our  calculations are striking. Although common $t$-$J$-type  models do not exhibit long-ranged DDW order,  a separate class of {\sl repulsive} Hamiltonians show robust long range DDW order even in the presence of {\sl substantial doping}.  These have their historical origin in a half-filled SO(5) invariant model on a
ladder \cite{Hanke}. At precise half-filling, it was shown to exhibit  DDW in its phase diagram \cite{John,MFS}. We shall show that SO(5) 
invariance is irrelevant  by considering coupling constants very far from the ``SO(5) parameters" and by substantially doping this model. The real reason for success with this class of models is that it straddles CDW and
DSC-like ($d_{x^2-y^2}$-superconductor) states (more precisely rung-singlet states), resulting in a local kinetic exchange  between them. This is much like the actual situation in the cuprates in which the DDW phase
is an intermediate regime between a multiplicity of complex charge ordered states and DSC \cite{CKN,John,MFS,Furusaki}.

We  label the  site of a ladder by ${\bf i}\equiv (r,l)$, where $r=1,\ldots,L$ is the rung index and $l=1,2$ is the leg
index. The current operator between any two sites $\bf i$ and $\bf j$, ${\cal J}_{{\bf i},{\bf j}}$,  is
\begin{equation}
{\cal J}_{{\bf i},{\bf j}} = i t \sum_{\sigma} \left( c^\dag_{{\bf i},\sigma}c_{{\bf j},\sigma} - c^\dag_{{\bf j}\sigma}c_{{\bf i},\sigma}\right),
\end{equation}
where $c^{\dagger}_{\bf i,\sigma}$ is the creation operator of a fermion with spin $\sigma$
at site $\bf i$. We set the lattice spacing to unity.

There are at least two convenient ways of probing DDW order. 
One approach is to measure the equal-time rung-rung current correlation function in the ground state
\begin{equation}
C(r,R)= \langle j_{\rm rung}(R+r/2) j_{\rm rung}(R-r/2) \rangle ,
\end{equation}
where $j_{\text{rung}}(r)={\cal J}_{(r,1),(r,2)}$. In order to minimize the effect of the boundaries
of a finite ladder, we should choose $R$ to be the location of the central rung, and we shall 
denote this correlation function as $C(r)$

An alternative approach is to break the time reversal symmetry explicitly by 
applying a source  $-h j_{\text{rung}}(1)$ on one end of the ladder and measure the current
induced  in the sample. The
source term for DDW is necessarily complex and a
complex DMRG program is needed, which is more demanding on memory 
and computer time. Nonetheless, we have used both methods for cross checks for every single case 
discussed in the present Letter. Of course, the sensitivity with respect to the magnitude of the source 
must be tested, and it must be made sure that it is not so large that we are probing some excited states.
We check this issue very carefully in all our numerical work. For both methods, we use a finite size algorithm, which is more reliable, 
performing sweeps to reach convergence \cite{DMRG}.

To orient ourselves we shall begin with the two-leg $t$-$J$ ladder, which is defined by the Hamiltonian:
\begin{equation}
H_{\rm tJ} = -t \sum_{\langle {\bf i},{\bf j} \rangle,\sigma}({\cal P}c^{\dagger}_{\bf i,\sigma} 
c_{\bf j,\sigma}{\cal P} + \text{h.c}.) \nonumber + J\sum_{\langle{\bf i},{\bf j}\rangle} ({\bf S}_{\bf i} \cdot{\bf S}_{\bf j} - \frac{ n_{\bf i} n_{\bf j}}{4}),
\end{equation}
where $n_{\bf i}$ is the total fermion occupation number at the site $\bf i$, and $\bf S_i$ is the spin 
operator at the corresponding site. ${\cal P}$
projects out doubly occupied sites and $\langle {\bf i},{\bf j}\rangle$ denotes 
pairs of nearest neighbor sites. The $t$-$J$ model is the simplest model that captures some aspects of the 
high temperature superconductors. Removal of electrons from the system is quantified by the 
doping parameter $\delta = 1- \langle n_i\rangle$. In
actual experiments, DSC is observed in the range $\delta = 0.05-0.25$; 
at small values of $\delta$ lies the pseudogap regime.
In this model, the DDW correlations decay exponentially with a correlation 
length $\xi = 3- 4$ \cite{Scalapino}. Additional next-nearest neighbor hopping, 
augmenting the model, may be supposed to suppress the competing CDW and DSC phases and thus
reveal DDW order, but our calculations and those of Ref. ~\cite{Tsutsui} do not support this idea. 
\begin{figure}
\centerline{\includegraphics[scale=0.4]{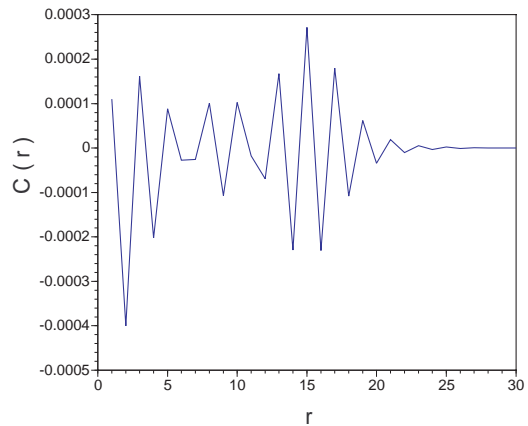}}
\caption{The correlation function $C(r)$ of the $t$-$J$-$V$-$V'$ model. 
The parameters are $J/t=0.4$, $V/t=3$, $V'/t=1$, and $\delta = 0.1$. The size of the 
ladder is 120x2. We kept 800 states and performed 42 sweeps. The vertical scale is chosen in units where t =1.}
\label{fig:Cofr}
\end{figure}

A more interesting model is
the $t$-$J$-$V$-$V'$ model, which is a $t$-$J$ model augmented by Coulomb repulsion
terms $V$ and $V'$ between nearest and next nearest neighbors:
\begin{equation}
H_{\rm tJVV'} = H_{\rm tJ} +V\sum_{\langle {\bf i},{\bf j} \rangle}n_{\bf i} n_{\bf j}+V'\sum_{\langle\langle {\bf i},{\bf j} \rangle\rangle}n_{\bf i} n_{\bf j} ,
\label{eq:tJVV}
\end{equation}
where $\langle\langle {\bf i},{\bf j} \rangle\rangle$ denotes next nearest neighbor pairs of sites. 
In order to reduce the effect of open boundaries, a  chemical potential term
$(V+V')n_{\bf i}$ is added to the boundary sites. 
A typical parameter set, given by $J/t=0.4$, $V/t=3$, $V'/t=1$, and $\delta= 0.1$, yields the current correlation function 
$C(r)$ shown in Fig.~\ref{fig:Cofr}.
\begin{figure}
\centerline{\includegraphics[scale=0.4]{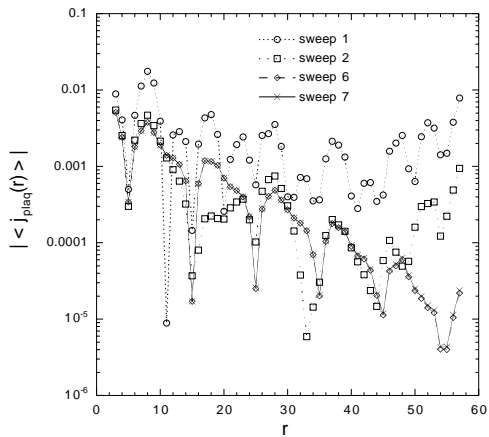}}
\caption{Absolute value of the current summed around a plaquette, $|\langle j_{\rm plaq}(r)\rangle|$, as induced by an edge current, 
$h=0.01t$, as a function of the
distance $r$ from the edge of a $60\times2$  ladder in the $t$-$J$-$V$-$V'$ model, with the parameters described 
in the text. Note the convergence as a function of sweeps. The number of states retained were 400 and the results were found to scale with current at this level.}
\label{fig:uli}
\end{figure}
 We observe 
what appears to be a bubble of DDW extending up to 20 rungs, and then a sinusoidally modulated 
exponential decay with $\xi \approx 3$. Thus, although we observe DDW
over a moderately long range, there is no macroscopic order. We have probed this model 
by the second method in which we induce a current by a source at the
edge. As shown in Fig. \ref{fig:uli},  the long range correlations that are observed at the
first infinite-size step disappear after  a few sweeps 
and converge after six sweeps to an exponential decay with a correlation length $\xi\approx10$. 
The moderately large bubble-like nature of $C(r)$
and the different values of $\xi$ in these two methods are strong reasons for suspecting 
proximity to a first order transition to the ordered DDW phase, as the boundary effects seem to nucleate this phase.

\begin{figure}
\centerline{\includegraphics[scale=0.4,angle=270]{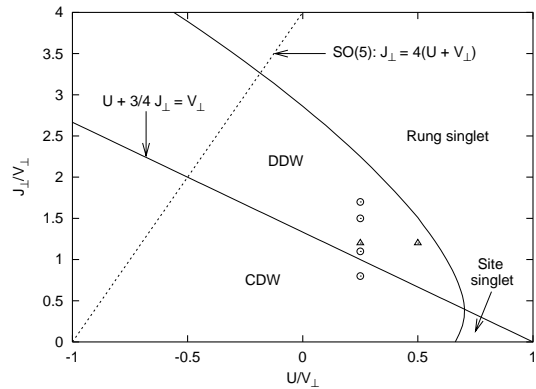}}
\caption{The weak coupling phase diagram at half filling for  $H_{\rm tJ_{\perp}UV_{\perp}}$ from a bosonization calculation. The open circles correspond to the parameters in Fig.~\ref{fig:4pct} and the open triangles to those in Fig.~\ref{fig:8pct}.}
\label{fig:rungpd}
\end{figure}

Finally, we shall consider  a different class of Hamiltonians, in which  a pair of electrons across a rung is given an 
internal structure, much like a molecule.  This is an interesting way 
of generating a set of low energy Hamiltonians \cite{Hanke} that are defined by
\begin{eqnarray}
&&H_{\rm tJ_{\perp}UV_{\perp}} = -t \sum_{\langle {\bf i},{\bf j} \rangle\sigma} (c^{\dagger}_{\bf i,\sigma} 
c_{\bf j,\sigma} + \text{h.c.})+\frac{U}{2} \sum_{{\bf i}}(n_{\bf i}-1)^2\nonumber \\
&+&\sum_{r}\left[J_{\perp}{\bf S}_{r,1}\cdot{\bf S}_{r,2}+V_{\perp}(n_{r,1}-1)(n_{r,2}-1)\right].
\label{eq:tJUV}
\end{eqnarray} 
 Note that there are no longer any projection operators 
$\cal P$, as in the previous examples and the problem can be treated for arbitrary interaction strength. 
In the half-filled case this Hamiltonian has a precise
SO(5) symmetry \cite{Hanke} when $J_{\perp}=4(U+V_{\perp})$. It also exhibits DDW order \cite{John,MFS}. 
But as soon as the system is doped, or the parameters are no longer finely tuned, there is no SO(5) symmetry. 
The weak coupling phase diagram at half-filling obtained from bosonization/RG, as shown in Fig.~\ref{fig:rungpd}, gives us some guidance as to where to look in our DMRG calculations. This phase diagram is essentially identical to that of Ref.~\cite{Furusaki}, except that we also show the regime for $U<0$.
Other than the DDW and CDW, there are two relevant states that can be adiabatically continued to resonating valence bond states \cite{RVB1} 
of the short-range variety \cite{RVB2}. These states are represented  by \cite{Troyer}
\begin{eqnarray}
|\mbox{rung-singlet}\rangle
\propto
\prod_r
\left[ \left| 
  \begin{array}{c}
     \uparrow \\ \downarrow
  \end{array}
\right\rangle_r
- 
\left| 
  \begin{array}{c}
     \downarrow \\ \uparrow
  \end{array}
\right\rangle_r
\right]
,
\label{eq:D-Mott}
\end{eqnarray}
and 
\begin{equation}
|\mbox{site-singlet}\rangle
\propto
\prod_r
\left[
\left| 
  \begin{array}{c}
     \uparrow\downarrow \\ - 
  \end{array}
\right\rangle_r
+
\left| 
  \begin{array}{c}
     - \\ \uparrow\downarrow
  \end{array}
\right\rangle_r
\right]
.
\label{eq:S-Mott}
\end{equation}
The DDW lies between the CDW and the rung-singlet phases. 

\begin{figure}
\centerline{\includegraphics[scale=0.4]{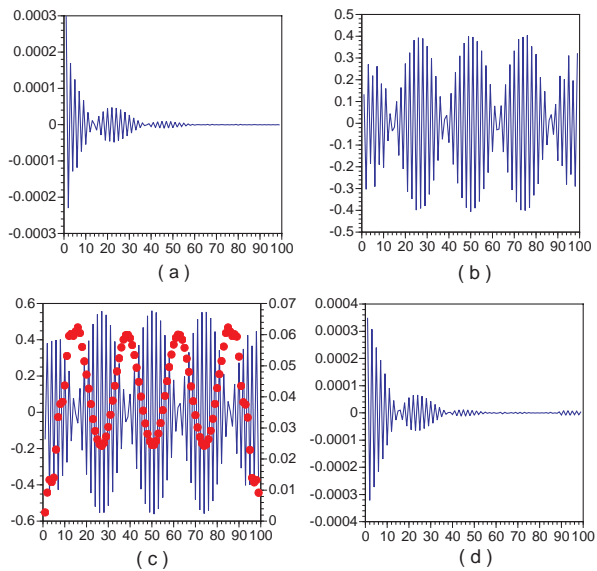}}
\caption{Rung current $j_{\rm rung}(r)$ as a function of the location of the rung $r$ in a $t$-$J_{\perp}$-$U$-$V_{\perp}$ model 
at 4\% doping on a 100x2 ladder, with parameters $U=0.25$, $t=V_{\perp}=1$ and an edge current of 0.0001. The sequence of figures correspond to (a) $J_{\perp}=0.8$, (b) $J_{\perp}=1.1$,
(c) $J_{\perp}=1.5$, and (d) $J_{\perp}=1.7$. In (c), we show the profile of the hole density depicted as solid dots corresponding to the scale on the right. We kept up to
400 states and performed up to 8 sweeps. Note the vast differences in the scales of the current strengths.}
\label{fig:4pct}
\end{figure}

\begin{figure}
\centerline{\includegraphics[scale=0.4]{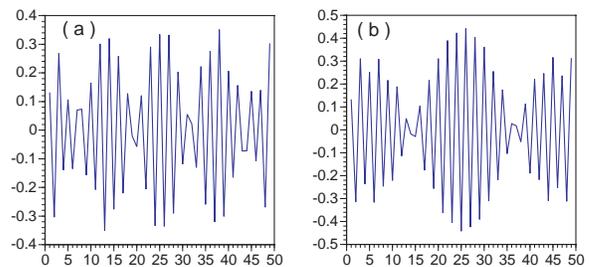}}
\caption{Rung current $j_{\text rung(r)}$ as a function of the location of the rung r in a $t$-$J_{\perp}$-$U$-$V_{\perp}$ model on a 50x2 ladder, 
with parameters $t=V_{\perp}=1$, $J_{\perp}=1.2 $ and an edge current of 0.0001. The sequence of figures correspond to (a) $U=0.25$ at 8\% doping, and (b) $U=0.5$ at 4\% doping. We kept up to 400 states and performed up to 8 sweeps.}
\label{fig:8pct}
\end{figure}

We have studied the Hamiltonian in Eq.~(\ref{eq:tJUV})
for a range of parameters and find long-range DDW order in the doped model, 
which has nothing to do with SO(5) symmetry. Nonetheless, the DDW phase is situated between the CDW and the rung-singlet phases, similar to the weak coupling bosonization results for half-filled ladders.  As a typical example, we have shown
in Fig.~\ref{fig:4pct} our results for the rung current as induced by an edge current of tiny magnitude 0.0001t. 
The parameters chosen were $U=0.25$, $t=V_{\perp}=1$, $J_{\perp}=0.8, 1.1,
1.5, 1.7$, and $\delta=0.04$. As a response, we see robust long-ranged DDW order in the middle of 
this range of $J_{\perp}$  with stripe like features where pairs of holes
reside; see, in particular, Fig.~\ref{fig:4pct}(c), where we also plot the hole density, and the coexistence with stripe order  
is especially evident from the antiphase domain wall structure \cite{stripe}.
The induced currents clearly alternate in sign and can be of order unity, in
units of $t$, even though the source current is infinitesimally small. For ladders of lengths 100, 150, and 200, and for parameters of Fig.~\ref{fig:4pct}(c), the current amplitudes are respectively 0.56, 0.53, and 0.53 respectively, consistent with long range order, though in a numerical calculation it is never possible to rule out a very slow decay.
We have studied the d-wave pairing correlations, and find only extremely rapid decay in ladders
that exhibit DDW long-range order. This is 
as expected from previous numerical work on the half-filled system \cite{MFS} where it 
was shown that the superconducting correlations decay exponentially in the DDW phase.  It is
also in accord with  the phase diagram in Fig.~\ref{fig:rungpd}, which shows that the region  with strong d-wave superconducting correlations, the ``rung-singlet'' phase,
is distinct from the DDW region.  More generally, bosonization leads to the prediction that the phase with strong DDW correlations has only short-ranged pairing correlations. For the case of Fig.~\ref{fig:4pct}(c),  we have also studied the low-lying
excitations about the ground state.  We find a robust spin gap, again in accord with general expectations.
In Fig.~\ref{fig:8pct}, we show the results for $\delta=0.08$ and for $U=0.5$. For sufficiently strong
doping, roughly between 10 to 20\%, DDW is suppressed for these sets of couplings. 
We have confirmed the existence of DDW on longer ladders, up to 200 rungs,
with current amplitudes decaying only slightly at the boundaries. 
It should be noted, however, that in two-leg ladders, there cannot be nodes in the excitation spectrum, as in two dimensions, 
which can, in principle, change the detailed nature of the doped phase. 

In summary,  we have shown that there are repulsive microscopic models that exhibit DDW order at finite doping, providing
added support for the identification of the pseudogap in the cuprates with this state. Our work also raises the real possibility
that  an appreciation of the complexity of many novel materials\cite{Coleman} may be  impossible
without these remarkable broken symmetries, and it is important to search for such complex quantum order in an even wider class of Hamiltonians.

We thank Chetan Nayak and Asle Sudb{\o} for stimulating discussions.
We acknowledge support from a Gerhard-Hess prize of the Deutsche Forschungsgemeinschaft (U. S.), the U. S. 
National Science Foundation, Grant No.  NSF-DMR-9971138 (S. C.) and Grant No.  NSF-DMR-9712391 (J. B. M), the Research Council of Norway, Grant. No.
142915/432 (J. O. F), and the  Swiss National Science Foundation (M. T.).

\end{document}